\documentclass[fleqn,10pt]{wlscirep}
\usepackage{amsmath,amsfonts,amssymb,graphics,graphicx,epsfig,color,times,bbm,subfigure,hyperref,keyval}

\title{Monogamy relation of multi-qubit systems for squared Tsallis-\emph{q} entanglement}
\author[1]{Guang-Ming Yuan}

\author[2,$^\ast$]{Wei Song}

\author[1,$^\dag$]{Ming Yang}

\author[2,$^\ddag$]{Da-Chuang Li}

\author[1]{Jun-Long Zhao}

\author[2]{Zhuo-Liang Cao}
\affil[1]{School of Physics and Material Science, Anhui University, Hefei, 230601, China}
\affil[2]{Institute for Quantum Control and Quantum Information, and School of Electronic and Information Engineering,
Hefei Normal University, Hefei 230601, China}

\affil[$^\ast$]{weis@hfnu.edu.cn}
\affil[$^\dag$]{mingyang@ahu.edu.cn}
\affil[$^\ddag$]{dachuangli@ustc.edu.cn}

\begin{abstract}
Tsallis-$q$ entanglement is a bipartite entanglement measure which
is the generalization of entanglement of formation for $q$ tending to 1. We first expand the range of $q$ for the analytic formula of Tsallis-\emph{q} entanglement. For $\frac{5-\sqrt{13}}{2} \leq \emph{q} \leq \frac{5+\sqrt{13}}{2}$, we prove the monogamy relation in terms of the squared
Tsallis-$q$ entanglement for an arbitrary multi-qubit systems. It is shown that the multipartite entanglement indicator based on squared Tsallis-$q$ entanglement still works
well even when the indicator based on the squared concurrence loses its efficacy. We also show that the $\mu$-th power of Tsallis-\emph{q} entanglement satisfies the monogamy or polygamy inequalities for any three-qubit state.
\end{abstract}

\begin{document}
\flushbottom
\maketitle

\section*{Introduction}
Quantum entanglement as a physics resource for quantum communication and quantum information processing has been the subject of many recent studies in recent years \cite{QE,qw,qr,qt,qy,qu,qi}. The study of quantum entanglement from various view points has been a very active area and has led to many interesting results. Monogamy of entanglement(MOE) \cite{MoE} is an interesting property discovered recently in the context of multi-qubit entanglement, which means that quantum entanglement cannot be shared freely in multi-qubit quantum systems. The bipartite monogamy inequality was first proposed and proved by Coffman, Kundu and Wootters(CKW) in a three-qubit system \cite{CKW}, and it is also named as CKW inequality:
\begin{equation}
C^{2}(\rho_{A|B C}) \geq C^{2}(\rho_{A B})+C^{2}(\rho_{B C}),
\end{equation}
where $C_{i j}^{2}$ is the squared of concurrence between the pair $i$ and $j$\cite{ABC}. Later, the monogamy inequality was generalized into various entanglement measures such as continuous-variable entanglement\cite{Adesso06,Hiroshima,Adesso07}, squashed
entanglement\cite{Koashi,Christandl,Yang}, entanglement negativity \cite{Ou,Kim09,He,Choi,Luo}, Tsallis-\emph{q} entanglement\cite{Kim10,Kim16}, and
R\'{e}nyi-$\alpha$ entanglement\cite{Kim10jpa,Cornelio10,Wang}. The applications of monogamy relation include many fields of physics such as characterizing the entanglement structure in multipartite
quantum systems\cite{Bai07pra,Bai08pra,Ou08,Jung,Eltschka,Ren,Cornelio,Regula,Kim14,Bai09,Zhu14pra,Zhu15pra,Bai13pra,Aguilar,Liu}, the security proof in quantum cryptography\cite{Masanes}, the frustration effects observed in condensed
matter physics\cite{PE}, and even black hole physics \cite{PQ,PW,PE,PR,PT,PY}. Originally, MOE was established in terms of the squared concurrence(SC).
Analogously, Bai \emph{et al} \cite{AB,Bai14pra} have proved that the
squared entanglement of formation(SEF) obeys the monogamy relation in arbitrary $N$-qubit mixed state. It should be noted that the entanglement of formation(EOF) itself does not satisfy the monogamy relation
even for three-qubit pure states. The new monogamy
relation in terms of SEF overcomes some flaws of the SC and can be used
to detect all genuine multipartite entanglement for $N$-qubit
systems.

On the other hand, Tsallis-$q$ entanglement is also a well-defined entanglement measure which
is the generalization of EOF. For $q$ tending
to 1, the Tsallis-$q$ entanglement converges to the EOF. A natural question
is whether the monogamy relation can be generalized to Tsallis-$q$ entanglement. In fact, Kim has derived a monogamy relation
in terms of Tsallis-$q$ entanglement\cite{Kim10}. However, the result in Ref\cite{Kim10} fails in including EOF as a special case and only holds for $2 \leq \emph{q} \leq 3$. In this paper we further
consider the monogamy relation in terms of the squared
Tsallis-$q$ entanglement(ST\emph{q}E). Firstly we expand the range of $q$ for the analytic formula of Tsallis-\emph{q} entanglement. Then we prove a monogamy inequality of multi-qubit systems in
terms of ST\emph{q}E in an arbitrary $N$-qubit
mixed state for $\frac{5-\sqrt{13}}{2} \leq \emph{q} \leq \frac{5+\sqrt{13}}{2}$, which covers the case of EOF as
a special case. Finally, we show that the $\mu$-th power of the Tsallis-\emph{q} entanglement satisfies the monogamy inequalities for three-qubit state.

\section*{Results}
\subsection*{Analytic formula of Tsallis-\emph{q} entanglement}

Firstly we recall the definition of Tsallis-\emph{q} entanglement introduced in Ref\cite{Kim10}. For a bipartite pure state$|\psi\rangle_{AB}$, the Tsallis-\emph{q} entanglement is defined as
{\begin{equation}
\emph{T}_{q}(|\psi\rangle_{AB}):= S_{q}\left ( \rho _{A} \right )= \frac{1}{q-1}\left ( 1-tr\rho _{A}^{q} \right ),
\end{equation}
for any $\emph{q} > 0$ and $\emph{q} \neq 1$, where $\rho_A=tr_B | \psi \rangle_{AB} \langle \psi |$ is the reduced density matrix by tracing over the  subsystem $B$. For the case when \emph{q} tends to 1, $\emph{T}_{q}\left ( \rho  \right )$ converges to the von Neumann entropy, that is
\begin{equation}
\lim_{q\rightarrow 1} \emph{T}_{q}\left ( \rho  \right )=-tr\rho \log \rho =S\left ( \rho  \right ).
\end{equation}

For a bipartite mixed state $\rho _{AB}$, Tsallis-\emph{q} entanglement is defined via the convex-roof extension
\begin{equation}
T_{q}\left ( \rho _{AB} \right ):=\min\sum_{i}p_{i}T_{q}(|\psi_{i}\rangle_{AB}),
\end{equation}
where the minimum is taken over all possible pure state decompositions of $\rho_{AB}=\sum_{i}p_{i} |\psi_{i}\rangle_{AB} \langle\psi_{i}|$.

In Ref\cite{Kim10}, Kim has proved an analytic relationship between Tsallis-\emph{q} entanglement and concurrence for $1\leq q \leq4$ as follows
\begin{equation}
 T_{q}(|\psi\rangle_{AB})=g_{q}(C(|\psi\rangle_{AB})),
\end{equation}
where the function $g_{q}(x)$ is defined as
\begin{equation}
g_{q}(x)=\frac{1}{q-1}\left[1-\left(\frac{1+\sqrt{1-x^2}}{2}\right)^q-\left(\frac{1-\sqrt{1-x^2}}{2}\right)^q\right],
\end{equation}

\begin{figure}[htb]
\includegraphics[scale=1.1]{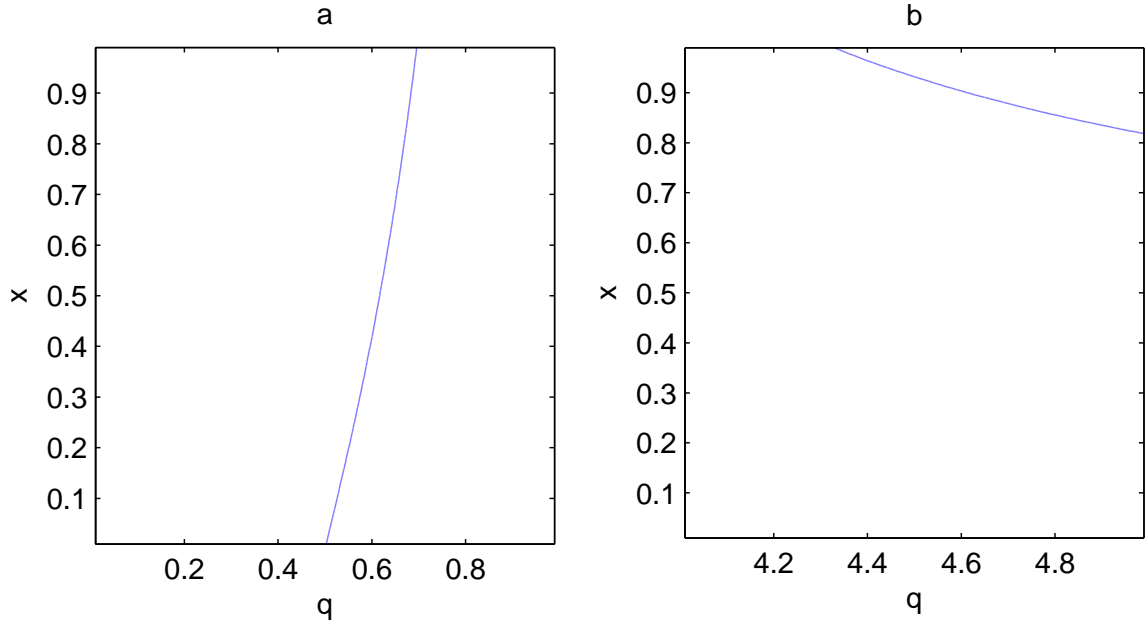}
\caption{The plot of the dependence of $x$ with $q$ which satisfies the equation $\frac{\partial^2 {T_{q}}}{\partial x^2}=0$ for (a)$q\in(0,1)$ and (b)$q\in(4,5)$ respectively.}\label{FIG1}
\end{figure}

According to the results in Ref\cite{Kim10}, the analytic formula in Eq.(5) holds for any $q$ such that $g_{q}(x)$ in Eq.(6) is monotonically increasing and convex. Next we shall generalize the range of $q$ when the function $g_{q}(x)$ is convex and monotonically increasing
with respect to $x$. The monotonicity and convexity of $g_{q}(x)$ follow from
the nonnegativity of its first and second derivatives. After a direct calculation, we find that the first derivative of $g_{q}(x)$ with respect to $x$ is always nonnegative for $q\geq0$\cite{Kim10}. Kim has also proved the nonnegative of the second-order derivative $g_{q}(x)$ for $1\leq q \leq4$.
We can further consider the second-order derivative of $g_{q}(x)$ beyond the region $1\leq q \leq4$. We first analyze the nonnegative region for the second-order derivative $g_{q}(x)$ for $q\in(0,1)$. Numerical calculation shows that under the condition $\partial^2 {T_{q}(C)}/\partial x^2=0$, the critical value of $x$ increases monotonically with the parameter $q$. In Fig.1(a), we plot the solution $(x,q)$ to this critical condition, where for each fixed $x$ there exists a value of $q$ such that the second-order derivative of $T_{q}(C)$ is zero. Because $x$ varying monotonically with $q$, we should only consider the condition $\partial^2 {T_{q}(C)}/\partial x^2=0$ in the limit $x\rightarrow1$. When $x = 1$, we have

\begin{equation}
\lim_{x \rightarrow 1}\frac{\partial^2 {T_{q}}}{\partial x^2}=-\frac{2^{1-q}(3-5q+q^2)}{3}\geq0,
\end{equation}
which gives the critical point $q_{c1}=\frac{5-\sqrt{13}}{2}\approx0.7$. When $q>q_{c1}$, the second-order $\partial^2 {T_{q}}/{\partial x^2}$ is always nonnegative. For $q\in(4,5)$, we find that the value of $x$ decreases monotonically with respect to $q$ as shown in Fig.1(b). In order
to determine the critical point we should only consider
the condition $\partial^2 {T_{q}}/\partial x^2=0$ in the limit $x\rightarrow1$. After direct calculation, we can obtain that the critical point $q_{c2}=\frac{5+\sqrt{13}}{2}\approx4.3$. When $q<q_{c2}$, the second-order $\partial^2 {T_{q}}/{\partial x^2}$ is always nonnegative. Combining with the previous results in Ref\cite{Kim10}, we get that the second derivative of $g_{q}(x)$ is always a nonnegative function for $\frac{5-\sqrt{13}}{2}\leq q \leq \frac{5+\sqrt{13}}{2}$. Thus we have shown that the analytic formula of Tsallis-\emph{q} entanglement in Eq.(5) holds for $\frac{5-\sqrt{13}}{2} \leq q \leq \frac{5+\sqrt{13}}{2}$.

\subsection*{Monogamy inequalities for ST$q$E in $N$-qubit systems.}
In the following we consider the monogamy properties of ST\emph{q}E. Using the results presented in Methods, we can prove the main result of this paper.

For an arbitrary \emph{N}-qubit mixed state $\rho_{A_1 A_2 \cdots A_n}$, the squared Tsallis-\emph{q} entanglement satisfies the monogamy relation
\begin{equation}
{T}_q^{2}(\rho_{A_1| A_2\cdots A_n}) \geq \sum_{i=2}^{n} T_q^{2}(\rho_{A_1 A_i}),
\end{equation}
where ${T}_q{(\rho_{A_1| A_2\cdots A_n})}$ quantifies the Tsallis-$q$ entanglement in the partition $A_1| A_2\cdots A_n$ and ${T}_q{(\rho_{A_1 A_i})}$ quantifies the one in two-qubit subsystem $A_1 A_i$ with the parameter $\frac{5-\sqrt{13}}{2}\leq \emph{q} \leq \frac{5-\sqrt{13}}{2}$.

For proving the above inequality, we first analyze an \emph{N}-qubit pure state $|\psi\rangle_{A_1 A_2 \cdots A_n}$. Under the partition $A_1| A_2\cdots A_n$, we have
\begin{eqnarray}
T_{q}^{2}(|\psi\rangle_{A_1| A_2\cdots A_n})  =  T_{q}^{2}[C_{A_1| A_2\cdots A_n}^{2}(|\psi\rangle)]
  \geq  T_{q}^{2}\left(\sum_{i=2}^{n} C^2_{A_1 A_i}\right)
  \geq  \sum_{i=2}^{n} T_{q}^{2}(\rho_{A_1 A_i}),\end{eqnarray}
where in the first inequality we have used the monogamy relation of squared concurrence $C_{A_1| A_2 \cdots A_n}^{2} \geq \sum_{i=2}^{n} C_{A_1 A_i}^{2}$ and the monotonically increasing property of $T_{q}^{2}(C^{2})$ which has been proved in Methods, and the second inequality is due to the convex property of $T_{q}^{2}(C^{2})$ (The details for proving the convexity property can be seen from Methods).

Next, we prove the monogamy relation for an \emph{N}-qubit mixed state $\rho_{A_1 A_2\cdots A_n}$. In this case, the formula of Tsallis-$q$ entanglement cannot be applied to $T_q(\rho_{A_1| A_2\cdots A_n})$ since the subsystem ${A_2\cdots A_n}$ is not a logic qubit in general. But we can still use the definition of Tsallis-$q$ entanglement in Eq.(4). Thus, we have

\begin{equation}
\emph{T}_{q}(\rho_{A_1| A_2\cdots A_n})=\min_{\{p_{i},|\psi_{i}\rangle\}} \sum p_{i} \emph{T}_{q}(|\psi_{i}\rangle_{A_1|A_2\cdots A_n}),
\end{equation}
where the minimum is taken over all possible pure state decompositions $\{p_{i},|\psi_{i}\rangle\}$ of the mixed state $\rho_{A_1| A_2\cdots A_n}$. Under the optimal decomposition $\{p_{j},|\psi_{j}\rangle_{A_1 |A_2\cdots A_n}\}$, we have
\begin{eqnarray}
T_{q}^{2}(\rho_{A_1| A_2\cdots A_n})& = & [\sum_{j} p_{j}T_{q}(|\psi_{j}\rangle_{A_1| A_2\cdots A_n})]^{2}
 = \{\sum_{j}p_{j}T_{q}[C_{A_1 |A_2\cdots A_n}(|\psi_{j}\rangle)]\}^{2} \nonumber\\
 &\geq&  \{T_{q}[\sum_{j}p_{j}C_{A_1| A_2\cdots A_n}(|\psi_{j}\rangle)]\}^{2}
 \geq  \{T_{q}[C_{A_1| A_2\cdots A_n}(\rho)]\}^{2}
 =  T_{q}^{2}[C_{A_1| A_2\cdots A_n}^{2}(\rho)]\nonumber\\
 &\geq& T_{q}^{2}[\sum_{i=2}^{n} C^{2}(\rho_{A_1 A_i})]
 \geq  \sum_{i=2}^{n} T_{q}^{2}[C^{2}(\rho_{A_1 A_i})]
 =  \sum_{i=2}^{n} T_{q}^{2}(\rho_{A_1 A_i}),
\end{eqnarray}
where in the second equality we have used the pure state formula of the Tsallis-\emph{q} entanglement and taken the $T_\emph{q}(C)$ as a function of the concurrence $C$ for $\frac{5-\sqrt{13}}{2} \leq \emph{q} \leq \frac{5+\sqrt{13}}{2}$; the third inequality is due to that $T_\emph{q}$ is a monotonically increasing and convex function of the concurrence for $\frac{5-\sqrt{13}}{2} \leq \emph{q} \leq \frac{5+\sqrt{13}}{2}$; the forth inequality is due to the convex property of concurrence for mixed state; and in the sixth and seventh inequalities we used the monotonically increasing and convex properties of $T_\emph{q}^2(C^2)$ as a function of the squared concurrence for $\frac{5-\sqrt{13}}{2} \leq \emph{q} \leq \frac{5+\sqrt{13}}{2}$ (The details for illustrating the property of ST$q$E can be seen from Methods). Thus we have completed the proof of the monogamy inequalities for ST$q$E in $N$-qubit systems.

As an application of the established monogamy relation in Eq.(8), we can construct the multipartite entanglement indicator $\tau_q(\rho)={T}_q^{2}(\rho_{A_1| A_2\cdots A_n}) - \sum_{i=2}^{n} T_q^{2}(\rho_{A_1 A_i})$ to detect the genuine multipartite entanglement. We consider a three-qubit pure state $|\psi(p)\rangle=\sqrt{p}|GHZ_3\rangle-\sqrt{1-p}|W_3\rangle$, which is the superposition of a GHZ state and a W state with $|GHZ_3\rangle=(|000\rangle+|111\rangle)/\sqrt{2}$ and $|W_3\rangle=(|001\rangle+|010\rangle+|100\rangle)/\sqrt{3}$. The three-tangle $\tau$ introduced in \cite{CKW} is defined as $\tau(|\psi(p)\rangle)=C_{A|BC}^2-C_{AB}^2-C_{AC}^2$. For the quantum state $|\psi(p)\rangle$, its three-tangle is $\tau(|\psi(p)\rangle)=p^2-8\sqrt{6}\sqrt{p(1-p)^3}/9$ which has two zero points at $p_1=0$ and $p_2\approx0.627$. On the other hand, we can directly calculate the value of $\tau_q(|\psi(p)\rangle)$ since the Tsallis-$q$ entanglement has an analytical formula for two-qubit quantum states. In Fig.2 we plot the three-tangle and the indicator $\tau_q$ for the order $q=0.8,1.1,1.4$. It is shown that the indicator $\tau_q$ is always positive for the different order $q$ in contrast to the three-tangle $\tau$ having two zero points. Thus we have shown that the indicator in terms of Tsallis-$q$ entanglement could detect the genuine entanglement in $|\psi(p)\rangle$ better than SC.

\begin{figure}[htb]
\includegraphics[scale=1]{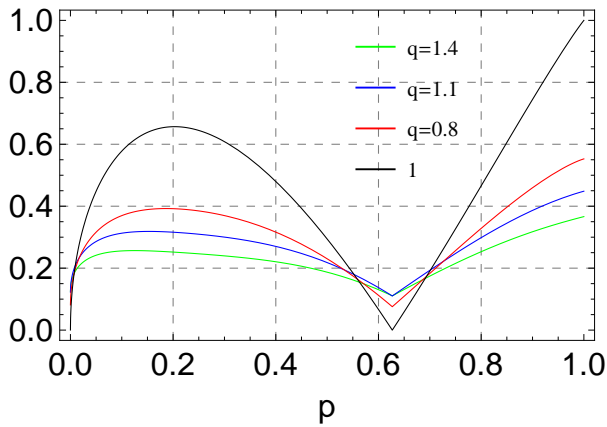}
\caption{The indicator $\tau_q$ for the superposition state $|\psi(p)\rangle$ with $q=0.8$(red line), $q=1.1$(blue line), and $q=1.4$(green line). We also plot the three-tangle of  $|\psi(p)\rangle$ with a black line.}\label{FIG2}
\end{figure}

\subsection*{Monogamy relation of the $\mu$-th
power of Tsallis-$q$ entanglement.}

Finally, besides the squared Tsallis-$q$ entanglement, we
can further consider the monogamy relation of the $\mu$-th
power of Tsallis-$q$ entanglement.

For any three-qubit state $\rho_{A_1 A_2 A_3}$, we can obtain
\begin{equation}
T_{q}^{\mu}(\rho_{A_1|A_2 A_3})\geq T_{q}^{\mu}(\rho_{A_1 A_2})+ T_{q}^{\mu}(\rho_{A_1 A_3}),
\end{equation}
for all $\frac{5-\sqrt{13}}{2} \leq \emph{q} \leq \frac{5+\sqrt{13}}{2}$, $\mu \geq 2$.

For proving Eq.(12), we consider the three-qubit case, according to the monogamy relation (8), we have
\begin{equation}
T_{q}^{2}(\rho_{A_1|A_2 A_3})\geq T_{q}^{2}(\rho_{A_1 A_2})+ T_{q}^{2}(\rho_{A_1 A_3}),
\end{equation}
for any three-qubit state $\rho_{A_1 A_2 A_3}$ with $\frac{5-\sqrt{13}}{2} \leq \emph{q} \leq \frac{5+\sqrt{13}}{2}$. Without loss of generality, assuming $T_{q}(\rho_{A_1 A_2})>T_{q}(\rho_{A_1 A_3})$, we can obtain
\begin{eqnarray}
T_{q}^{\mu}(\rho_{A_1| A_2 A_3})& \geq & (T_{q}^{2}(\rho_{A_1 A_2})+T_{q}^{2}(\rho_{A_1 A_3}))^{\frac{\mu}{2}}
 =  T_{q}^{\mu}(\rho_{A_1 A_2})\left( 1 + \frac{T_{q}^{2}( \rho _{A_1 A_3})}{T_{q}^{2}( \rho _{A_1 A_2})}\right )^{\frac{\mu}{2}}  \nonumber\\
& \geq & T_{q}^{\mu}(\rho_{A_1 A_2})\left(1+\left(\frac{T_{q}^{2}(\rho_{A_1 A_3})}{T_{q}^{2}(\rho_{A_1 A_2})}\right)^{\frac{\mu}{2}}\right)
 =  T_{q}^{\mu}(\rho_{A_1 A_2})+T_{q}^{\mu}(\rho_{A_1 A_3}),
\end{eqnarray}
where the second inequality comes from the property $(1+x)^{t}\geq1+x^{t}$ for $x \leq 1$, $t \geq 1$. If $T_{q}(\rho_{A_1 A_2})=0$ or $T_{q}(\rho_{A_1 A_3})=0$, the inequality obviously holds.

Similarly, we have the following polygamy inequalities. For any three-qubit $\rho_{A_1 A_2 A_3}$, we have
\begin{equation}
T_{q}^{\mu}(\rho_{A_1|A_2 A_3})\leq T_{q}^{\mu}(\rho_{A_1 A_2})+ T_{q}^{\mu}(\rho_{A_1 A_3}),
\end{equation}
for all $\frac{5-\sqrt{13}}{2} \leq \emph{q} \leq \frac{5+\sqrt{13}}{2}$, $\mu \leq 0$.

For any three-qubit state $\rho_{A_1 A_2 A_3}$ with $\frac{5-\sqrt{13}}{2} \leq \emph{q} \leq \frac{5+\sqrt{13}}{2}$, we have

\begin{eqnarray}
T_{q}^{\mu}(\rho_{A_1| A_2 A_3})& \leq & (T_{q}^{2}(\rho_{A_1 A_2})+T_{q}^{2}(\rho_{A_1 A_3}))^{\frac{\mu}{2}}
 =  T_{q}^{\mu}(\rho_{A_1 A_2})\left( 1 + \frac{T_{q}^{2}( \rho _{A_1 A_3})}{T_{q}^{2}( \rho _{A_1 A_2})}\right )^{\frac{\mu}{2}}  \nonumber\\
& < & T_{q}^{\mu}(\rho_{A_1 A_2})\left(1+\left(\frac{T_{q}^{2}(\rho_{A_1 A_3})}{T_{q}^{2}(\rho_{A_1 A_2})}\right)^{\frac{\mu}{2}}\right)
 =  T_{q}^{\mu}(\rho_{A_1 A_2})+T_{q}^{\mu}(\rho_{A_1 A_3}),
\end{eqnarray}
where in the second inequality we have used the inequality $(1+x)^{t} < 1+x^{t}$ for $x > 0$, $t \leq 0$.

\section*{Discussion}
In this paper we have generalized the analytic formula of Tsallis-\emph{q} entanglement to the region $\frac{5-\sqrt{13}}{2} \leq \emph{q} \leq \frac{5+\sqrt{13}}{2}$. Then we proved the monogamy relation in
terms of ST\emph{q}E for an arbitrary multi-qubit systems, which include previous result in terms of EOF as a special case. Based on the monogamy properties of Tsallis-\emph{q} entanglement, we have shown that the corresponding indicator can work well even when the indicator based on the squared concurrence loses its efficacy. In addition, we considered the monogamy or polygamy relation of the $\mu$-th power of Tsallis-\emph{q} entanglement. One distinct advantage of our result is that infinitely
many inequalities parameterized by $q$ provides greater flexibility than previous monogamy relation in terms of EOF.

\section*{Methods}
\subsection*{$T_{q}^{2}(C^2)$ is a monotonically-increasing function of the squared concurrence $C^{2}$ for all $q\geq0$.}

Notice that Eq.(5) can also be written as
\begin{equation}
 T_{q}(|\psi\rangle_{AB})=f_{q}(C^2(|\psi\rangle_{AB})),
\end{equation}
where the function $f_{q}(x)$ is defined as
\begin{equation}
f_{q}(x)=\frac{1}{q-1}\left[1-\left(\frac{1+\sqrt{1-x}}{2}\right)^q-\left(\frac{1-\sqrt{1-x}}{2}\right)^q\right].
\end{equation}

 The squared Tsallis-$q$ entanglement is a monotonically increasing function of $C^{2}$ if the first-order derivative $\partial{T_{q}^{2}(C^2)}/\partial x>0$ with $x = C^2$. By direct calculation, we have,

 \begin{equation}
 \frac{\partial {T_{q}^{2}(C^2)}}{\partial x}=2L(1-2^{-q}M^{q}-2^{-q}N^{q})\left[\frac{{2^{-1-q}q(M^{q-1}-N^{q-1})}}{\sqrt{1-x}}\right],
 \end{equation}
which is always nonnegative on $0 \leq x \leq 1$ for all $q \geq0 $, where $L=1/(q-1)^2$, $M=1+\sqrt{1-x}$, $N=1-\sqrt{1-x}$, and the equality holds only at the boundary. Thus we get that $T_{q}^{2}$ is a monotonically increasing function of $x$ with $x = C^2$.

\subsection*{$T_{q}^{2}(C^2)$ is a convex function of the squared concurrence $C^{2}$ for $\frac{5-\sqrt{13}}{2} \leq q \leq \frac{5+\sqrt{13}}{2}$.}

\begin{figure}[htb]
\includegraphics[scale=1.1]{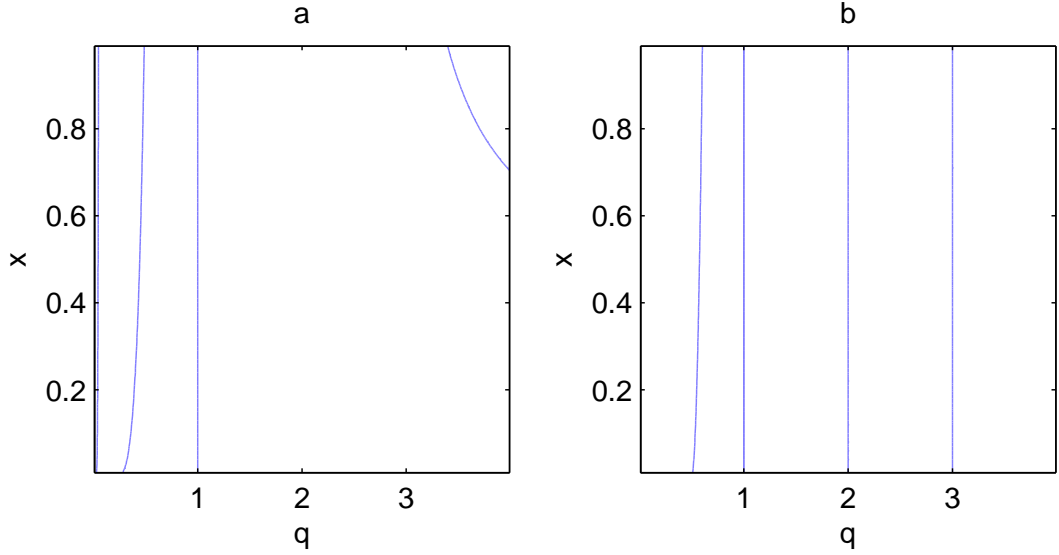}
\caption{The plot of the dependence of $x$ with $q$ which satisfies the equation (a)$\frac{\partial F_{q}}{\partial q}=0$ and (b)$\frac{\partial F_{q}}{\partial x}=0$ respectively.}\label{FIG3}
\end{figure}

The convex property of the squared concurrence is satisfied if the second-order derivative $\partial^2 {T_{q}^{2}(C^2)}/\partial x^2=\partial^2 {f_{q}^{2}(C^2)}/\partial x^2>0$ with $x = C^2$. We first define a function $F_{q} :=\partial^2 {[(q-1)^2T_{q}^{2}(C^2)]}/\partial x^2$ on the domain $D=\{(x,q)|0 \leq x \leq 1,1 \leq q \leq 4\}$, then the
nonnegativity of the second-order derivative $T_{q}^{2}$
can be guaranteed by the nonnegativity of $F_{q}$ since it varies with
$\partial^2 {T_{q}^{2}(C^2)}/\partial x^2$ by a positive constant. After some deduction,
we have

\begin{eqnarray}
F_{q} & & =
 \left\{2 \left( 1-2^{-q}M^q-2^{-q}N^q \right)\right. \left[\frac{2^{-2-q}q(M^{q -1}-N^{q-1})}{(1-x)^{3/2}}-\frac{2^{-2-q}(q-1)q(M^{q-2}+N^{q-2})}{1-x}\right]\nonumber\\
& &\left. + 2\left[\frac{{2^{-1-q}q(M^{q-1}-N^{q-1})}}{\sqrt{1-x}}\right]^{2}\right\}.
\end{eqnarray}

In order to prove the nonnegativity of $F_{q}$, it is suffice to consider its maximum or minimum values on the domain $D$. The critical points of $F_{q}$ satisfy the condition

\begin{equation}
\nabla F_{q}=\left(\frac{\partial F_{q}}{\partial x},\frac{\partial F_{q}}{\partial q}\right )=0.
\end{equation}

\begin{figure}[htb]
\includegraphics[scale=1.1]{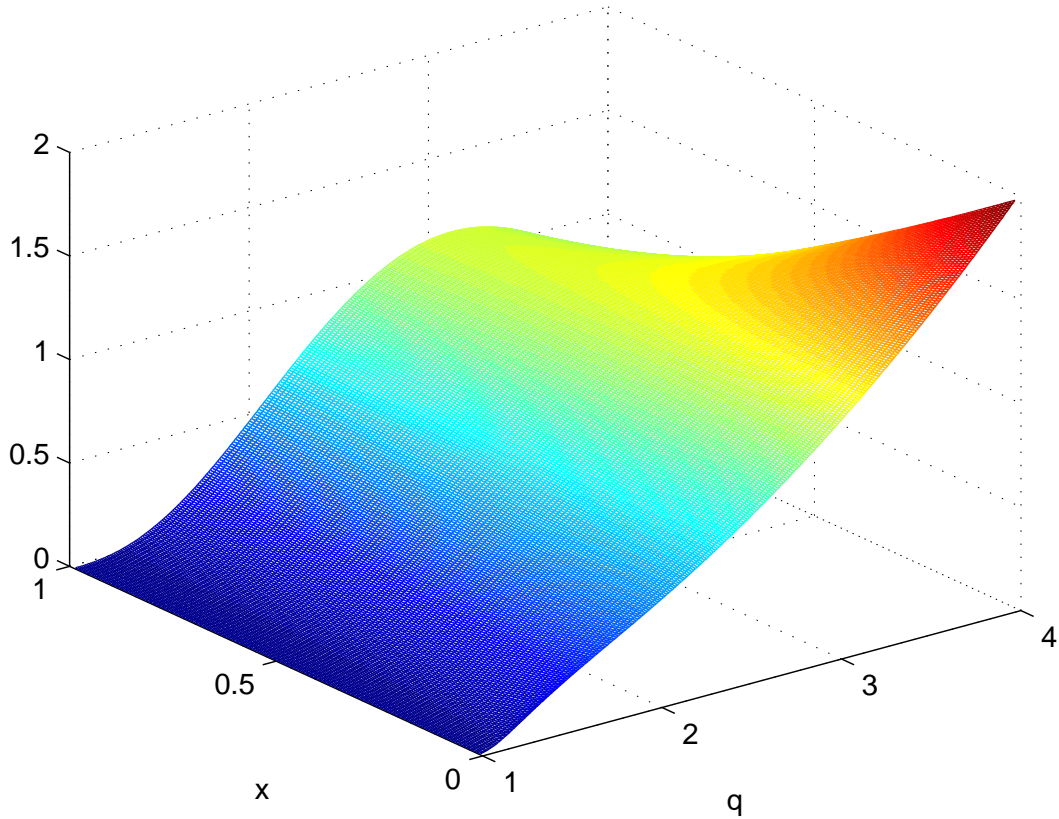}
\caption{$F_{q}$ is plotted as a function of $x$ and $\emph{q}$ for $0 \leq x \leq 1$,$1 \leq \emph{q} \leq 4$ }\label{FIG4}
\end{figure}

In Fig.3(a) and (b),we have plotted the value of $x$ and $q$ which satisfies the equation $\partial F_{q}/\partial q=0$ and $\partial F_{q}/\partial x=0$ respectively. Combining the results in Fig.3(a) and (b), we find that the solution of the above equation is $q=1$ which is one of the boundary of domain $D$. To ensure the nonnegative of $F_{q}$, we should only consider
the other two cases on the boundary of $F_{q}$, i.e., $x=0$ and $x=1$.

For the case $x=0$,
\begin{equation}
\lim_{x \rightarrow 0} F_{q}=2^{-1-2q}q(2^q-2)(q-1),
\end{equation}
which is always nonnegative in the region $\emph{q}\in(1,4)$.

For the case when $x=1$,
\begin{equation}
\lim_{x \rightarrow 1} F_{q}=\frac{4^{-q}(1-q)q[6(2^q -2)+(16-5\times2^q)q+(2^q-8)q^2]}{3},
\end{equation}
 where Eq.(23) is always nonnegative for $q=1$ and $q=4$, and the first-order derivative of Eq.(23) increases first and then decreases for $1\leq q \leq4$. Thus we prove that Eq.(23) is nonnegative in the region $1\leq q \leq4$. Notice that $F_{q}$ has no critical points in the interior of $D$, we conclude that $F_{q}$ is always nonnegative for $1\leq q \leq4$. The nonnegative of the $F_{q}$ is also plotted in FIG.4.

\begin{figure}[htb]
\includegraphics[scale=1.1]{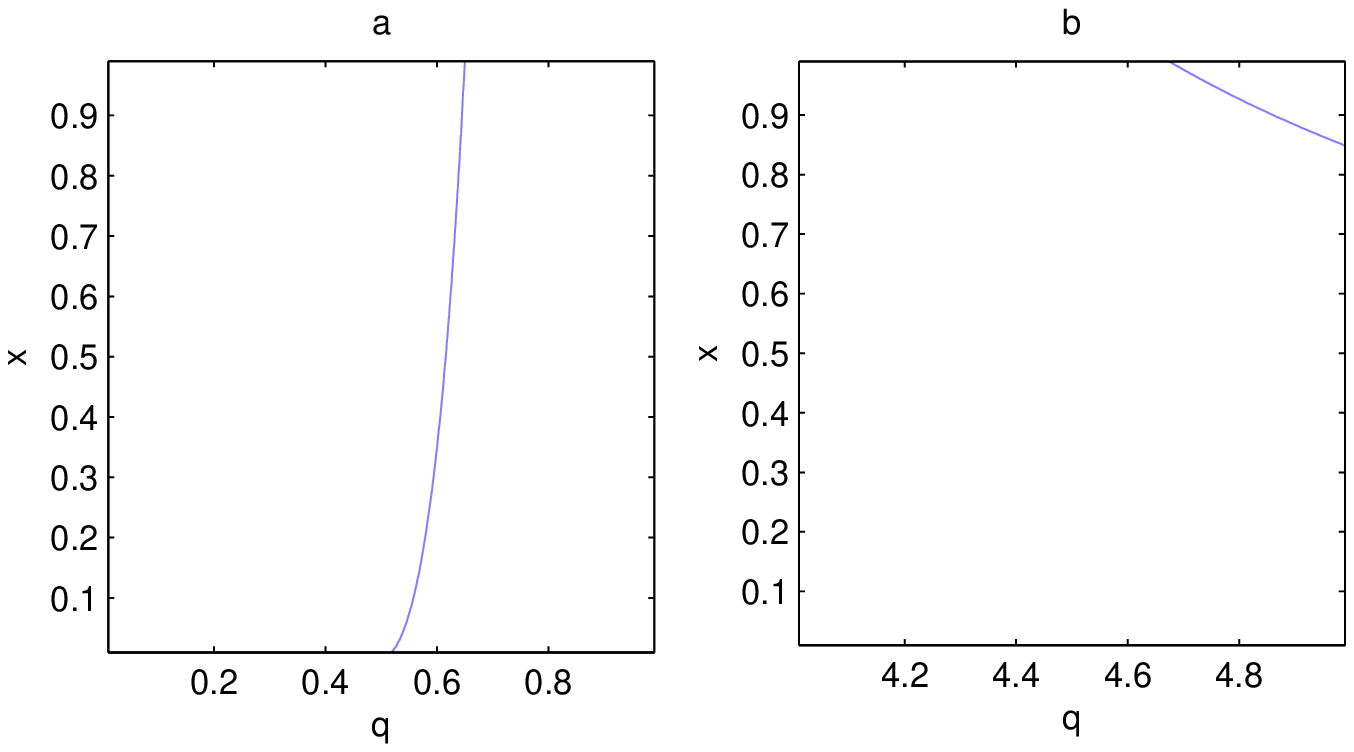}
\caption{The plot of the dependence of $x$ with $q$ using the equation $\frac{\partial^2 {T_{q}^2}}{\partial x^2}=0$ for (a)$q\in(0,1)$ and (b)$q\in(4,5)$ respectively.}\label{FIG5}
\end{figure}

Furthermore,we can consider the nonnegative region for the second-order derivative $\partial^2 {T_{q}^2}/\partial x^2$ when $q$ ranges in $(0,1)$. Under the condition $\partial^2 {T_{q}^2}/\partial x^2=0$, we find that the critical value of $x$ increases monotonically with the parameter $q\in(0,1)$. In Fig.5(a), we plot the solution $(x,q)$ to the critical condition $\partial^2 {T_{q}^2}/\partial x^2=0$ where for each fixed $x$ there exists a value of $q$ such that the second-order derivative of $T_{q}^2$ is zero. We should only consider the condition $\partial^2 {T_{q}^2}/\partial x^2\geq0$ in the limit $x\rightarrow1$. In this case, we have
\begin{equation}
\lim_{x \rightarrow 1}\frac{\partial^2 {T_{q}^2}}{\partial x^2}=-\frac{4^{-q}q[6(2^q -2)+(16-5\times2^q)q+(2^q-8)q^2]}{3(q-1)}\geq0,
\end{equation}
which gives the critical point $q_{c3}\approx0.65$. When $q\geq q_{c3}$, the second-order $\partial^2 {T_{q}^2}/\partial x^2$ is always positive. Similarly, we can also analyze the nonnegative region for the second-order derivative $\partial^2 {T_{q}^2}/\partial x^2$ when $q$ ranges in $(4,5)$. In Fig.5(b), it is shown that the critical value of
$x$ decreases monotonically along with the parameter $q\in(4,5)$, and the critical point $q_{c4}\approx4.65$. When $q\leq q_{c4}$, the second-order $\partial^2 {T_{q}^2}/\partial x^2$ is always positive. Notice that
the analytical formula of $T_{q}$ is established only for $\frac{5-\sqrt{13}}{2} \leq q\leq \frac{5+\sqrt{13}}{2}$, we conclude that the second-order derivative $\partial^2 {T_{q}^2}/\partial x^2$ is positive for $\frac{5-\sqrt{13}}{2} \leq q\leq \frac{5+\sqrt{13}}{2}$ which completes the proof of the convexity property of $T_{q}^{2}(C^2)$ with the squared concurrence $C^{2}$ for $\frac{5-\sqrt{13}}{2} \leq q \leq \frac{5+\sqrt{13}}{2}$.

\section*{Acknowledgements}
This work is supported by the National Natural Science Foundation of China (NSFC) under Grants No.11374085, No.11274010, No.11204061; the Anhui Provincial Natural Science Foundation under Grant No.1408085MA16; the Anhui Provincial Candidates for academic and
technical leaders Foundation under Grant No.2015H052; the
discipline top-notch talents Foundation and the Excellent Young
Talents Support Plan of Anhui Provincial Universities.

\section*{Author contributions statement}

G. M. Yuan and W. Song carried out the calculations. W. Song, M. Yang and D. C. Li conceived the idea. All authors contributed to the interpretation of the results and the writing of the manuscript. All authors reviewed the manuscript.

\section*{Additional information}

\textbf{Competing financial interests:} The author declares no competing financial interests.

\end{document}